\documentstyle[preprint,tighten,aps,psfig]{revtex}

\begin{document}

\preprint{TTP97-54, hep-ph/9712222}

\title{\boldmath Two-loop QCD corrections to the heavy quark pair
production cross section in $e^+e^-$ annihilation near the threshold}

\author{Andrzej Czarnecki}
\address{Physics Department, Brookhaven National Laboratory\\
Upton, NY 11973}

\author{Kirill Melnikov}
\address{Institut f\"ur Theoretische Teilchenphysik, 
Universit\"at Karlsruhe\\
D-76128 Karlsruhe, Germany}
\maketitle

\begin{abstract}
We present the ${\cal O}(\alpha _s ^2)$ corrections to the cross
section for the reaction $e^+e^- \to \gamma ^* \to Q\bar Q$ in the
energy region close to the threshold. We assume that the energy of the
reaction is such that both the perturbative expansion in the strong
coupling constant $\alpha_s$ and expansion in the velocity $\beta$ of
the heavy quarks can be used.  We explicitly obtain terms ${\cal
O}(\alpha _s^2/\beta^2, \alpha _s^2 /\beta,\alpha_s^2)$ in the
relative correction to the threshold cross section.  Using the ideas
of asymptotic expansions, we demonstrate how an expansion of Feynman
diagrams in the threshold region is constructed.  From this analysis
we obtain a matching relation between the vector current in full QCD
and the quark--antiquark current in NRQCD at leading order in $1/m$
and the second order in the strong coupling constant.
\end{abstract}

\vspace*{0.2cm}

Theoretical predictions for the cross section of the reaction $e^+e^-
\to Q\bar Q$ in the energy region close to the $Q\bar Q$ threshold are
of considerable interest for various phenomena.  They are important
for determinations of the $b$ and $c$ quark masses, as well as the
value of the QCD coupling constant $\alpha _s (\mu)$, with $\mu \sim
1-2$ GeV, if one uses the sum rule approach for $\Upsilon$ and
$J/\psi$ hadrons proposed in \cite {Voloshin:1995sf}; achievement of
the ${\cal O}(\alpha_s ^2)$ accuracy is considered very important for
these quantities.  One should also mention the ongoing efforts to
determine the decay rates of the heavy quarkonia to leptons with the
${\cal O}(\alpha _s ^2)$ accuracy \cite{Bodwin:1995jh}.  Also, for the
future $e^+e^-$ or $\mu^+\mu^-$ colliders one is considering precision
measurements of the top quark properties by studying its threshold
production region.  It is well known that for these and other
applications, where the threshold region is of interest, the fixed
order perturbative calculations break down and a resummation of the
terms singular at threshold is mandatory.  In the leading order, such
resummation yields the well known Sommerfeld--Sakharov factor for the
threshold production cross section.  Aiming at the ${\cal O}(\alpha _s
^2)$ accuracy for the threshold cross section, the first thing to be
calculated is the perturbative expression for the cross section at the
order ${\cal O}(\alpha _s ^2)$ in the energy region where $\alpha_s
\ll \beta \ll 1$, where $\beta$ denotes the quark velocity, $\beta =
\sqrt {1-4m^2/s}$, and $s$ is the total energy squared 
\cite{Hoang1,Hoang:1997sj}.

From the technical point of view, the problem of perturbative
calculations in the threshold region has never received much
attention.  On the one hand, it is clear that a small parameter in
which a useful expansion could in principle be constructed --- $\beta$
--- is there; on the other hand, it has not been quite clear how to
use this parameter systematically in order to obtain a significant
simplification of the Feynman integrals which should be calculated. On
the conceptual level, the recognition of the existence of small and
large scales in the threshold problems was formulated as the
Non--Relativistic QCD (NRQCD), an effective field theory which should
be used to describe physics in the threshold region \cite
{Bodwin:1995jh}.  In this framework the corrections which originate
from the scales $k \sim m$ can be incorporated through the so--called
matching procedure. This requires calculations in the
full QCD, in order to provide the matching conditions.  It is expected
that since the infrared behaviors of both QCD and NRQCD are the same,
the matching calculations could be considerably simplified.  To the
best of our knowledge, however, it has not yet been demonstrated, how
the matching calculations in QCD/NRQCD should be organized in order to
achieve such simplification in practice.

For practical reasons it is useful to be able to use the smallness of
the relative velocity of the quarks on the level of individual Feynman
diagrams, i.e.\ to formulate a prescription which operates with
diagrams and subgraphs, rather than with composite operators and
effective field theories.  The advantages of such approach are its
transparency and better control over the calculation.  Indeed, similar
approaches in other kinematic situations have recently permitted to
complete many previously impossible calculations, both in QCD and in
the electroweak theory (see for example
\cite{CKM96,Czarnecki:1997hc,Chetyrkin:1996ia}).

In this paper we consider  a variant of the asymptotic expansions
which can be used in the threshold region.  It allows to construct an
expansion in $\beta$ of a given Feynman diagram.  For the resulting
Feynman integrals one can construct algorithms which permit their
calculation in any order in $\beta$, and can be encoded in a symbolic
manipulation language.

An approach of this type has recently been discussed in
\cite{Beneke:1997zp}.  Although its correctness has not been proven,
its construction is analogous to the well established asymptotic
expansions \cite{Smirnov90,Chetyrkin:1988zz,Chetyrkin:1988cu}. 

Using this approach we calculate the ${\cal O}(\alpha_s ^2)$
correction to the cross section of the reaction $e^+e^- \to Q\bar Q$,
up to terms of order ${\cal O}(\beta^0)$ relative to the Born cross
section.  In the result the terms with a color factor $C_F^2$, 
as well as the corrections induced by vacuum polarization insertions 
due to heavy and light fermions
are identical to the previously obtained results in the Abelian gauge
theory \cite{Hoang:1997sj,Hoang:1995ex}.
This provides a non-trivial test of the
asymptotic expansion method applied in the threshold region.  The
non-abelian terms presented below are a new result. 

We begin with introducing some notations.
The cross section of the reaction $e^+e^- \to \gamma ^* \to \bar Q Q$
is written as
\begin{equation}
\sigma _{e^+e^- \to \bar Q Q} = \sigma ^{(0)} \left [
1 + C_F \left (\frac {\alpha _s}{\pi} \right ) \Delta ^{(1)} 
+  C_F \left (\frac {\alpha _s}{\pi} \right )^2 \Delta ^{(2)} \right ],
\end{equation}
where
\begin{equation} 
\sigma ^{(0)} (s) = \frac {4}{3}\frac {\alpha ^2}{s}\;N_c\;e_Q^2 
\frac {\beta(3-\beta^2)}{2},
\qquad
\beta = \sqrt {1-\frac {4m^2}{s}},
\end{equation}
$s$ is the total energy squared of the reaction, $m$ is the mass of
the heavy quark $Q$, $N_c = 3$ is the number of colors and $e_Q$ is
the charge of the quark $Q$ in units of the electric charge.  We also
always use $\alpha_s \equiv \alpha _s(m)$ in the $\overline {MS}$
scheme and the on--shell renormalization for the heavy quark
propagators.

\begin{figure} 
\hspace*{-36mm}
\begin{minipage}{16.cm}
\vspace*{3mm}
\[
\mbox{
\begin{tabular}{ccc}
\psfig{figure=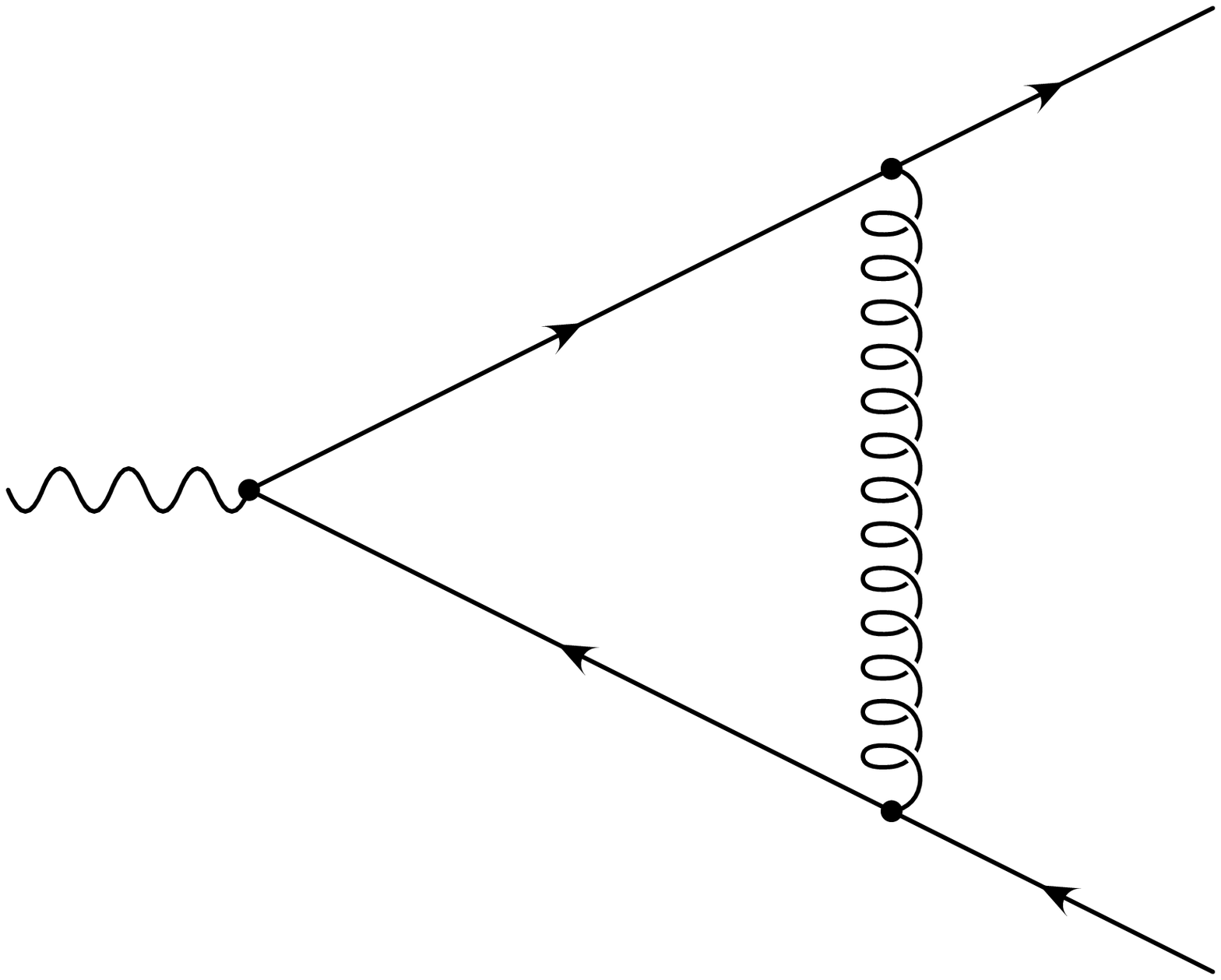,width=23mm,bbllx=210pt,bblly=410pt,%
bburx=630pt,bbury=550pt} 
&\hspace*{4mm}
\psfig{figure=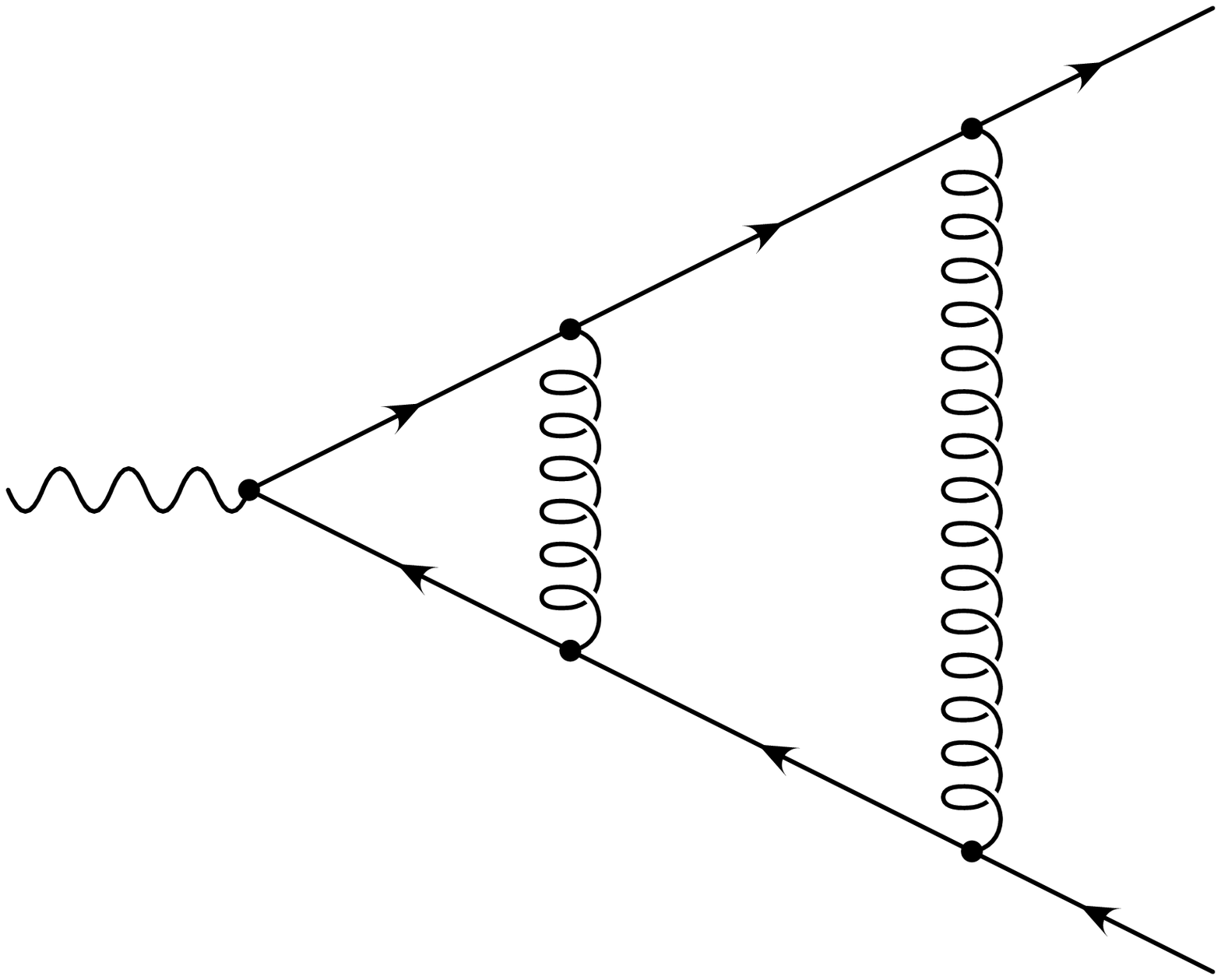,width=23mm,bbllx=210pt,bblly=410pt,%
bburx=630pt,bbury=550pt}
&\hspace*{4mm}
\psfig{figure=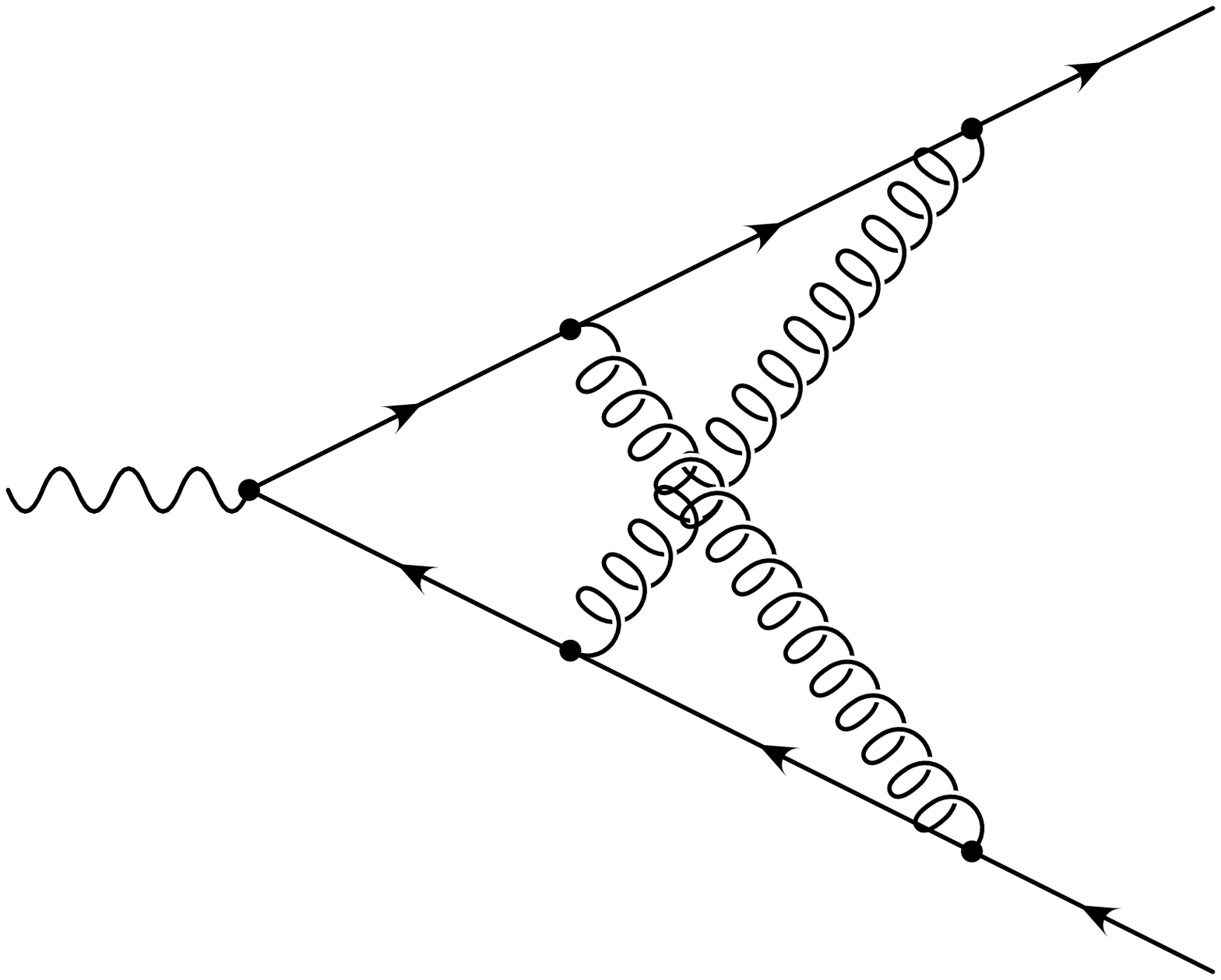,width=23mm,bbllx=210pt,bblly=410pt,%
bburx=630pt,bbury=550pt}
\\[13mm]
\hspace*{-13mm}(a) & \hspace*{1mm}(b) & \hspace*{1mm}(c) 
\\[7mm]
\psfig{figure=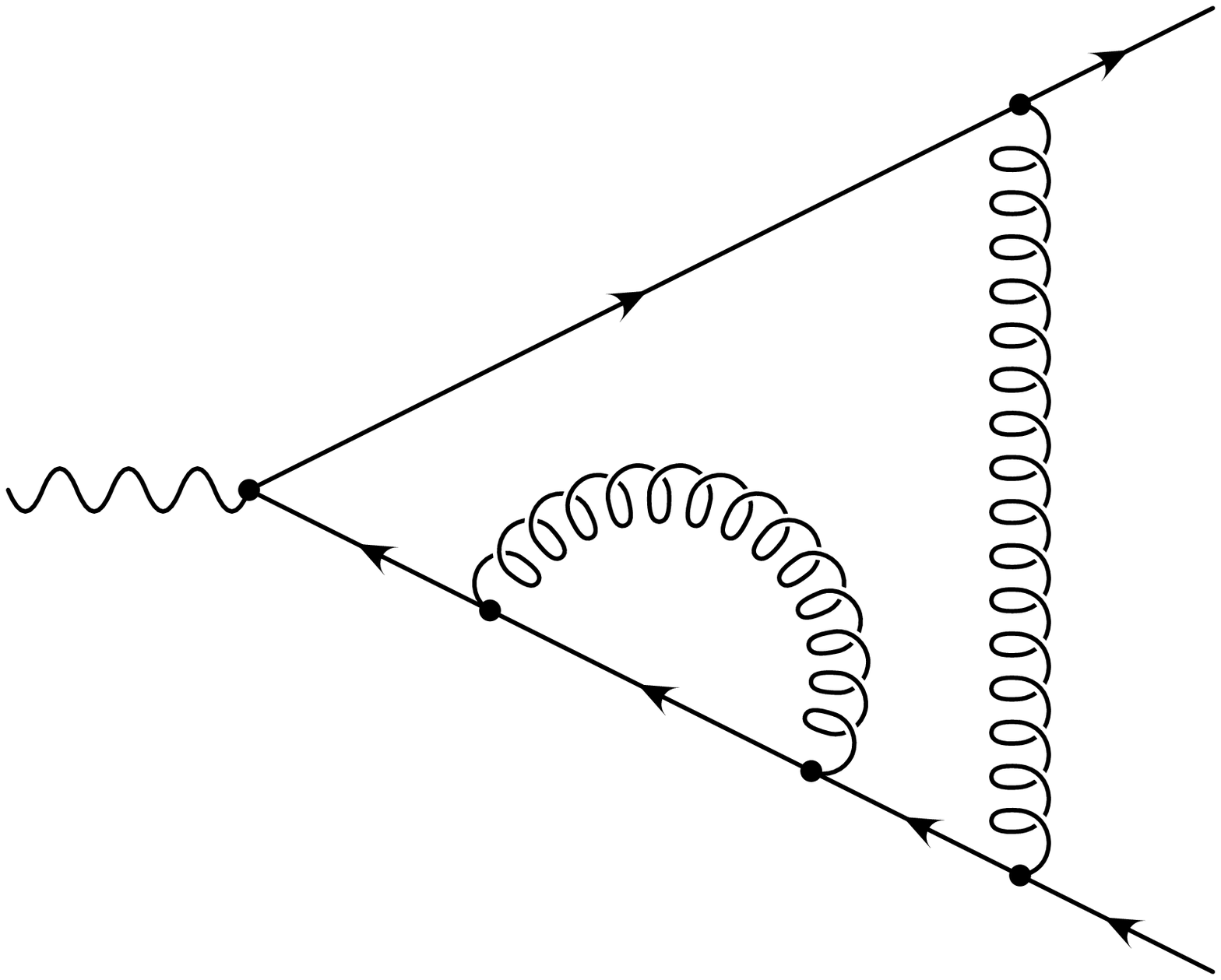,width=23mm,bbllx=210pt,bblly=410pt,%
bburx=630pt,bbury=550pt} 
&\hspace*{4mm}
\psfig{figure=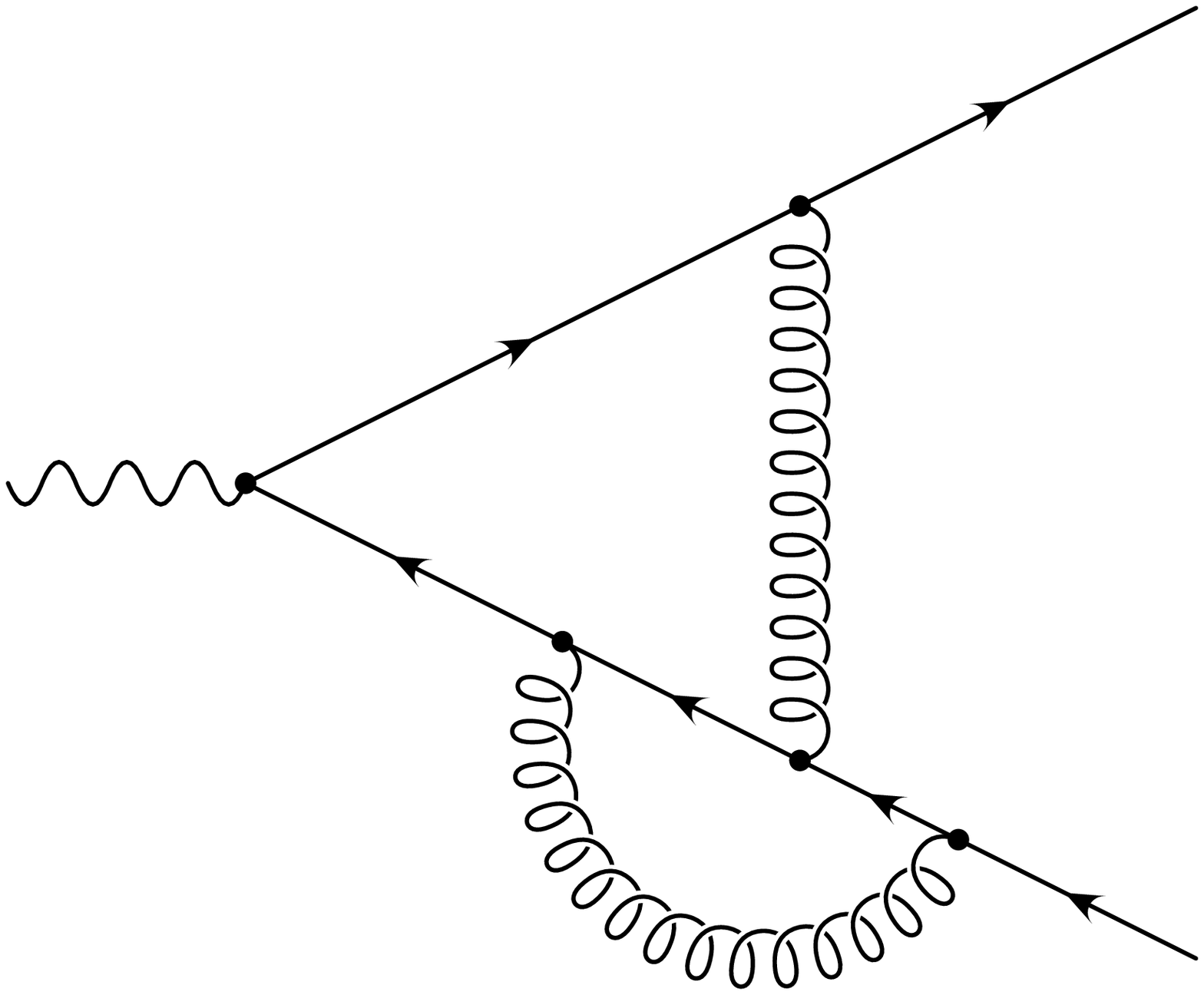,width=23mm,bbllx=210pt,bblly=410pt,%
bburx=630pt,bbury=550pt}
&\hspace*{4mm}
\psfig{figure=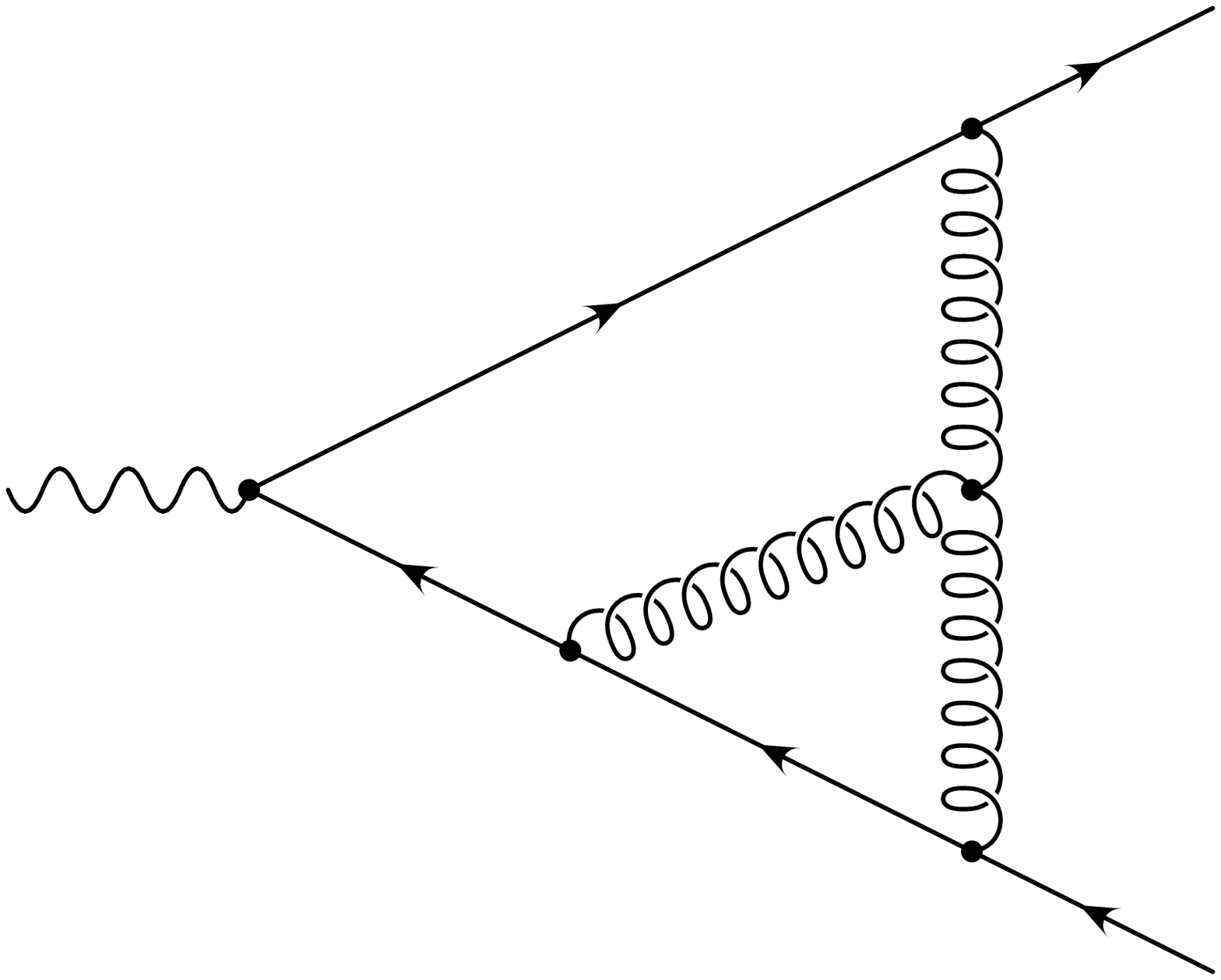,width=23mm,bbllx=210pt,bblly=410pt,%
bburx=630pt,bbury=550pt}
\\[13mm]
\hspace*{-13mm}(d) & \hspace*{1mm}(e) & \hspace*{1mm}(f) 
\\[7mm]
\psfig{figure=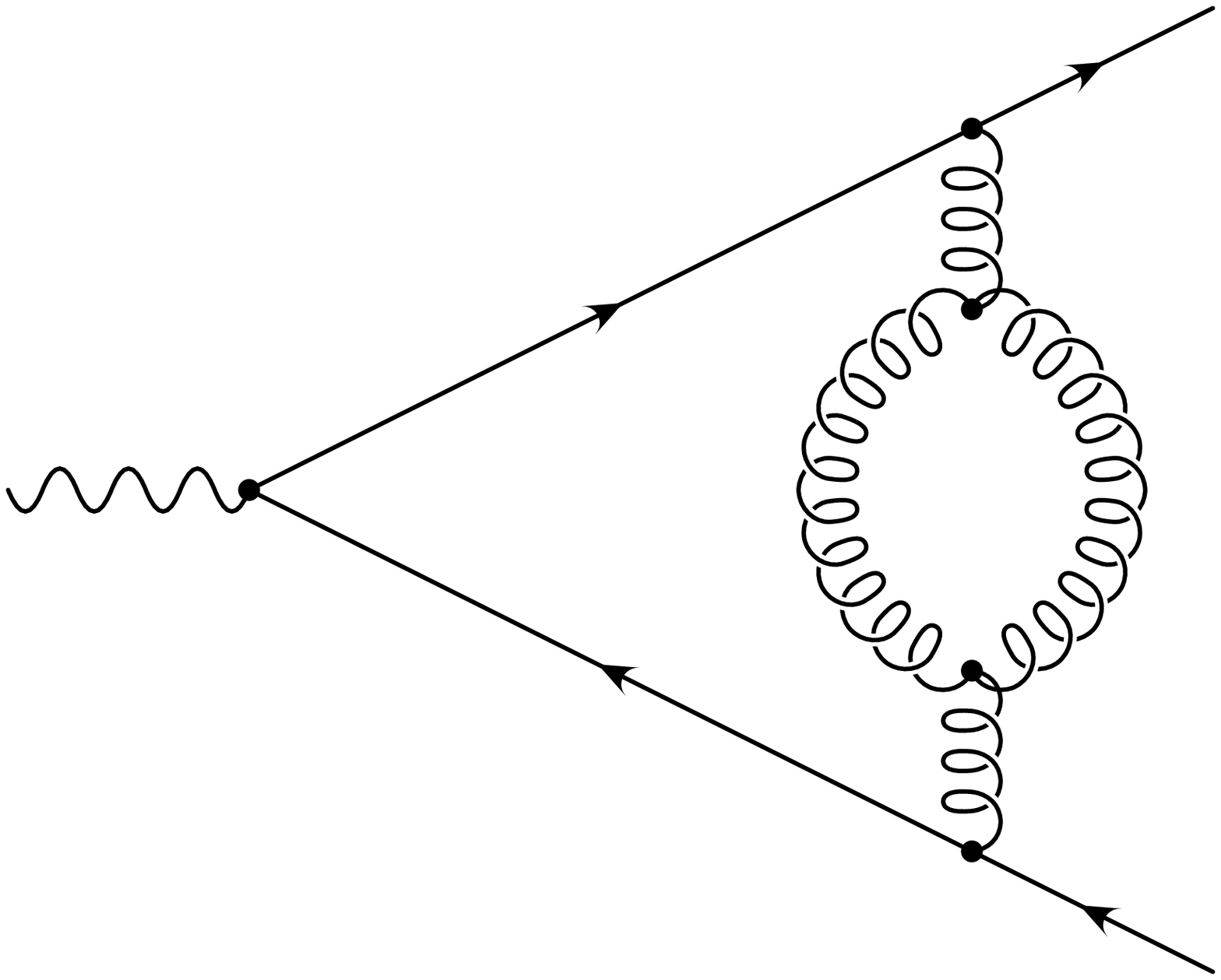,width=23mm,bbllx=210pt,bblly=410pt,%
bburx=630pt,bbury=550pt} 
&\hspace*{4mm}
\psfig{figure=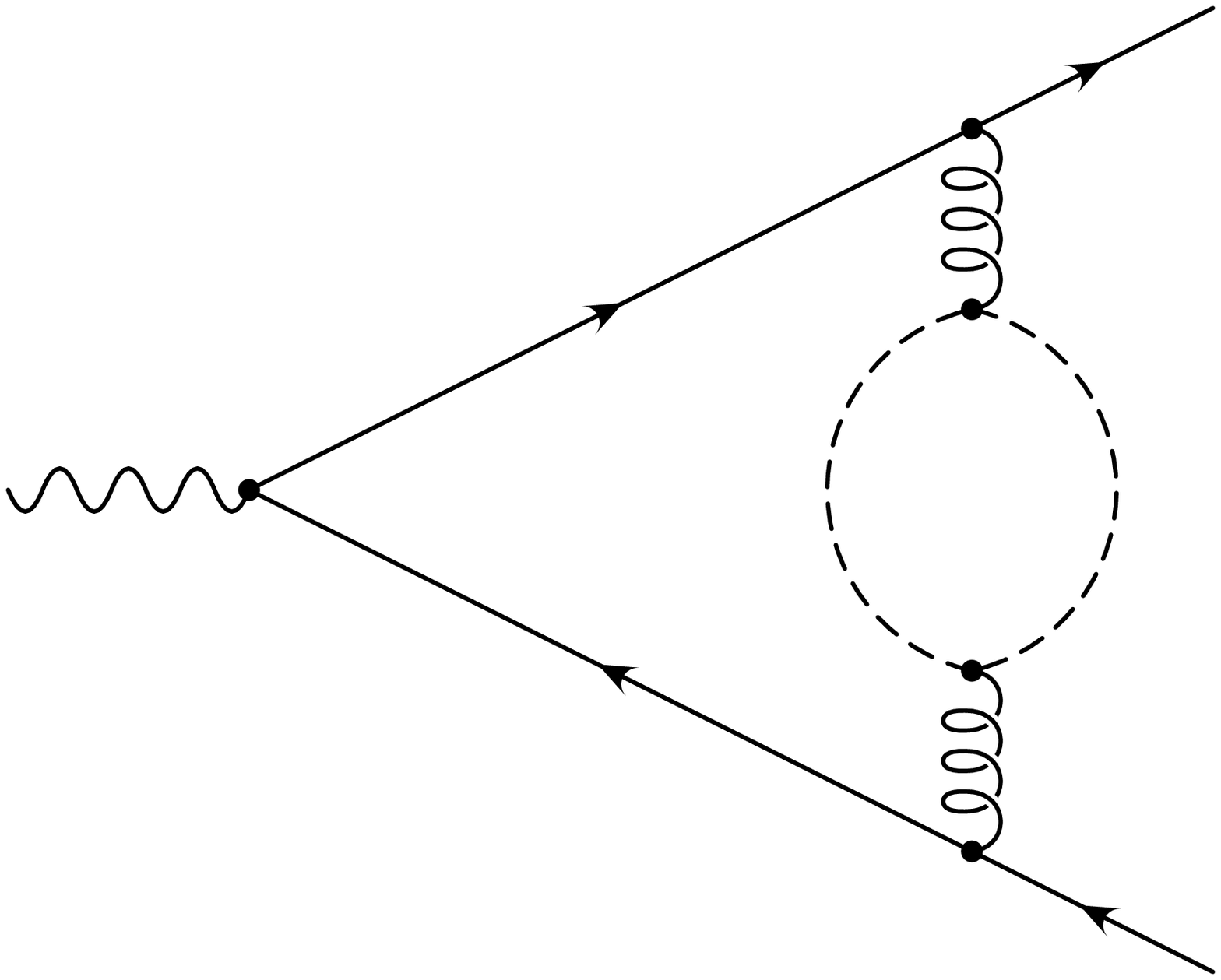,width=23mm,bbllx=210pt,bblly=410pt,%
bburx=630pt,bbury=550pt}
&\hspace*{4mm}
\psfig{figure=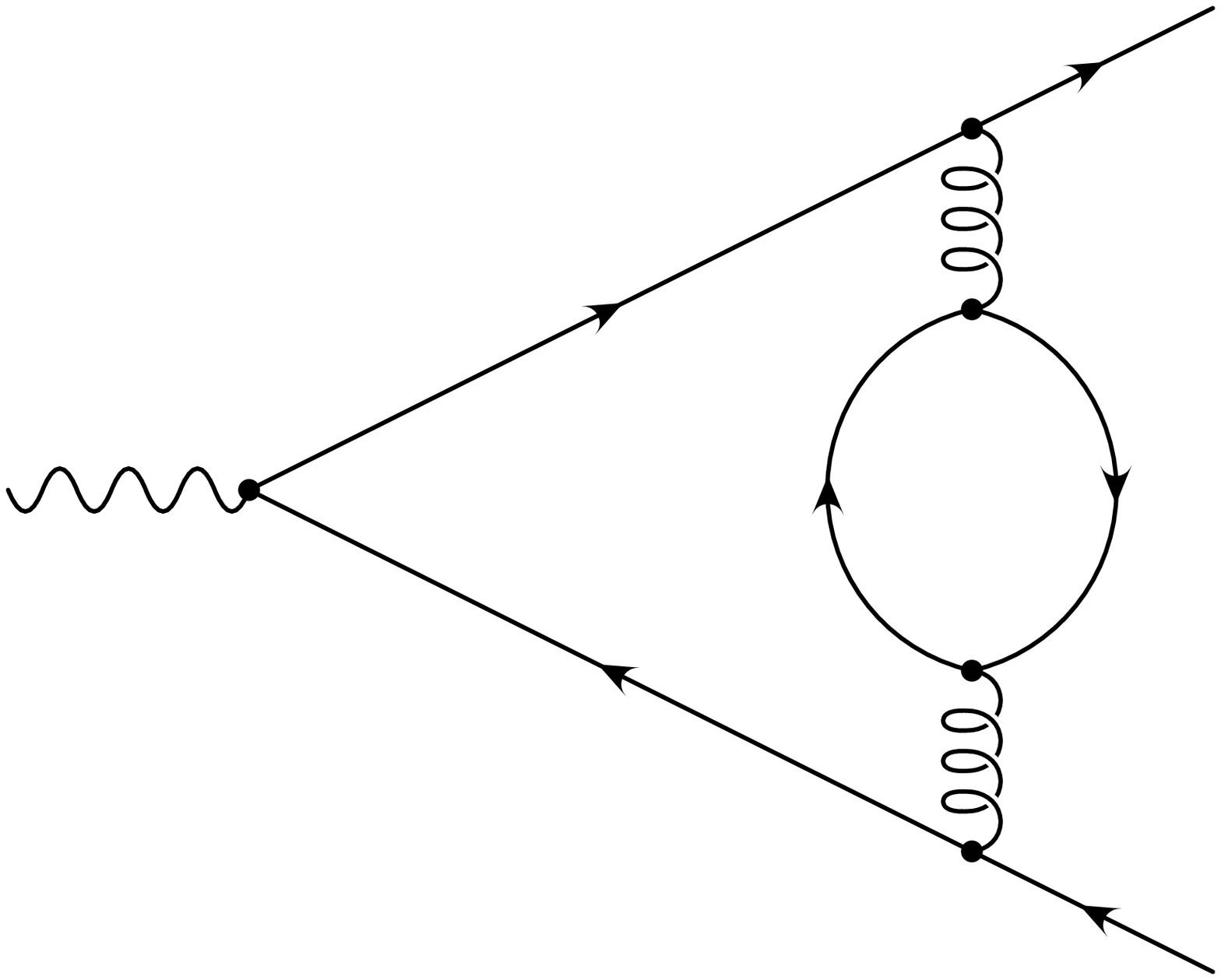,width=23mm,bbllx=210pt,bblly=410pt,%
bburx=630pt,bbury=550pt}
\\[13mm]
\hspace*{-13mm}(g) & \hspace*{1mm}(h) & \hspace*{1mm}(i) 
\\[4mm]
\end{tabular}}
\]
\end{minipage}
\caption{Two-loop  QCD corrections to the quark pair production; (a)
is the one loop vertex correction; (b-i) are the two loop gluonic and
fermionic corrections.}
\label{fig:vert}
\end{figure}

The terms $\Delta ^{(1)}$ and $\Delta ^{(2)}$ represent, respectively,
the one-- and two--loop corrections.  Vertex corrections which
contribute at these orders are shown in Fig.~\ref{fig:vert}.  The term
$\Delta ^{(1)}$ is known in exact form as a function of $\beta$.  Near
the threshold we are interested in the expressions for both 
$\Delta^{(1,2)}$ up to and including 
terms ${\cal O}(\beta ^0)$\footnote{The term ${\cal O}(\beta)$ in 
$\Delta ^{(1)}$ can also be used for the matching on non--relativistic
resummed cross--section. This term represents a kinematical
correction and leads, essentially, to the replacement
$\pi^2/(2\beta) \to \pi^2(1+\beta^2)/(2\beta)$, 
see \cite{Hoang1,Hoang:1997sj}.}.  We note,
that up to this order in $\beta$, there is no need to consider
radiation of real gluons in the process of interest.  With this
restriction the expression for $\Delta ^{(1)}$ reads
\begin{equation}
\Delta ^{(1)} = \frac {\pi^2}{2\beta} - 4 + {\cal O}(\beta).
\label {1}
\end{equation}
The two terms in this expression are known to be of rather different
origins.  The term proportional to the inverse power of $\beta$ is the
so--called Coulomb correction, while the second term is known to be a
hard correction.  Let us demonstrate how these corrections could be
calculated using asymptotic expansions.

In this approach for each diagram one has to identify those regions of
the integration momenta which can give non-vanishing contributions at
the given order in $\beta$.  In the calculations close to threshold,
one should distinguish four different regions (the following
description applies to the center of mass frame of the $Q \bar Q$
pair)  \cite {Beneke:1997zp}:
\begin{enumerate}
\item hard
region, where the momenta of the quanta are of the order of $k \sim
m$,
\item soft region, where $k_0 \sim |{\bf k}| \sim m\beta$,
\item potential region, where $k_0 \sim m\beta^2$ and $|{\bf k}| \sim
m\beta$,
\item
ultrasoft region $k \sim m\beta^2$.
\end{enumerate}
One should construct all possible subgraphs of a given graph assigning
the above labels to the lines in all possible combinations.  After
such assignment, the routing of the momenta should satisfy the
``scale conservation.''  Namely, two ultrasoft lines, for instance,
can not produce a potential line. On the contrary, two potential
lines can produce an ultrasoft line and so on.

To calculate  $\Delta^{(1)}$ to necessary order in $\beta$, one has
to consider the real part of the one--loop correction to the vertex
$\gamma ^* Q \bar Q$. In this diagram only potential and hard regions
contribute. The other two regions generate massless tadpoles which
vanish in dimensional regularization, which we use throughout this
calculation.  The potential subgraph gives rise to a finite
contribution $\pi^2/(2\beta)$. The hard subgraph (which to the order
${\cal O}(\beta ^{0})$ is obtained by simply considering the vertex
correction at the point $s=4m^2$ in dimensional regularization) gives
the constant term $-4$, if combined with the ultraviolet
renormalization of the Born cross section.

To see the importance of this classification we recall the following
fact.  The above result for $\Delta ^{(1)}$ is valid in the region
$\beta \gg \alpha _s$, where perturbative calculations are still
justified. When one approaches the region of small $\beta$, one should
perform a resummation of all terms of the form $ \alpha
_s^n/\beta^n$. Such resummation results in the well known
Sommerfeld--Sakharov factor, which is the modulus square of the
fermion wave function at the origin, when the Coulomb interaction
between nonrelativistic $Q$ and $\bar Q$ is taken into account
exactly.  It is also believed that   part of the 
subleading terms of the form
$\alpha _s^{n+1}/\beta^n$ can be resummed by multiplying the
Sommerfeld--Sakharov factor by the so--called hard correction:
\begin{eqnarray}
\sigma &=& \sigma ^{(0)} \left (1 - 4 C_F \frac {\alpha_s}{\pi} \right )
|\Psi (0) |^2,
\nonumber \\
|\Psi (0) |^2 &=& \frac {z}{1-\exp (-z)},
\qquad z = \frac {C_F \alpha_s \pi}{\beta}.
\end{eqnarray}
It is interesting that this hard renormalization constant for the
Sommerfeld--Sakharov factor, which has been known already for a long
time, is the same as the contribution of the hard subgraph of the
one--loop vertex correction and the ultraviolet renormalization. This
is of course not accidental.  If the approach based on asymptotic
expansions in the threshold calculations is to be trusted, it is fairly
clear, that the above formula is correct and the one--loop hard
correction does indeed provide the renormalization of the Coulomb
ladder to all orders in the coupling constant. A simplest way to see
this is to say that the contribution of the hard subgraph provides a
renormalization of the NRQCD vector current in order to match it on
the complete vector current in QCD (see a more detailed discussion
below).

One can hope that the knowledge of the contributions coming from hard
subgraphs at order ${\cal O}(\alpha _s ^2)$ would permit a
determination of the two--loop renormalization of the Coulomb ladder.

We now turn to the consideration of the second order correction,
described by the term $\Delta ^{(2)}$.  It is convenient to decompose
it into terms proportional to various SU(3) color factors (in SU(3)
$C_F=4/3$, $C_A=3$, $T_R=1/2$, and $N_{L,H}$ are the numbers of quark
flavors of mass 0 and $m$, respectively):
\begin{equation}
\Delta ^{(2)} = C_F~\Delta^{(2)}_A +C_A~\Delta ^{(2)}_{N\!A} 
 + N_LT_R~ \Delta ^{(2)}_L + N_H T_R~\Delta ^{(2)}_H.
\end{equation}
The only unknown term in this expression is $\Delta ^{(2)}_{N\!A}$,
which arises in the non--abelian theory.  $\Delta^{(2)}_{A,H,L}$ are
the same in the abelian and non--abelian theory. In the framework of
QED $\Delta^{(2)}_{A}$ were obtained in \cite {Hoang:1997sj} in the
threshold region.  The terms which describe the contribution of both
massless and massive 
fermions $\Delta ^{(2)}_{L,H}$ were calculated in an analytical
form for arbitrary $\beta$ in \cite {Hoang:1995ex}.  
Also, we would like
to mention, that the leading ${\cal O}(\alpha _s ^2/\beta)$ terms in
$\Delta ^{(2)}_{N\!A}$ were predicted in \cite {CKS}
using the observation that similar terms in $\Delta^{(2)}_{L}$ are
directly related to the contribution of light fermions to the
Coulomb--like QCD interaction potential between heavy quark and
anti--quark.   These terms can be also extracted from
the expression for the cross section for 
$e^+e^- \to Q\bar Q$, which includes the resummation of the terms 
$\alpha _s ^{(n+1)}/\beta^n$. 
Our calculation confirms  these results and allows a
determination of the ${\cal O}(\beta^0)$ terms in $\Delta
^{(2)}_{N\!A}$, which is the main new result of the present paper.

Various contributions to $\Delta ^{(2)}$ are:
\begin{eqnarray}
\Delta ^{(2)} _A &=& 
\frac {\pi ^4}{12\beta^2} -2 \frac {\pi^2}{\beta}
+ \frac {\pi^4}{6} 
\nonumber \\ &&
+ 
 \pi^2 \left ( - \frac {35}{18} - \frac {2}{3}\log \beta + \frac
 {4}{3}\log 2 \right ) +  \frac {39}{4} - \zeta _3, 
\\
\Delta ^{(2)} _{N\!A} &=& 
\frac {\pi^2}{\beta}
\left( \frac {31}{72} - \frac {11}{12}\log 2\beta  \right )
\nonumber \\ &&
+ \pi ^2 \left ( \frac {179}{72} - \log \beta - 
\frac {8}{3}\log 2 \right )  - \frac {151}{36} - \frac {13}{2}\zeta _3,
\\
\Delta ^{(2)} _L &=&  
\frac {\pi^2}{\beta} \left (\frac {1}{3} \log 2\beta -\frac {5}{18}
\right ) + \frac {11}{9},
\\
\Delta ^{(2)} _H &=& \frac {44}{9} - \frac {4\pi^2}{9}.
\end{eqnarray}
The terms $\Delta ^{(2)} _{A,L,H}$ coincide with the results obtained
in QED.  
The  non-abelian piece $\Delta ^{(2)} _{N\!A}$ was studied
numerically  in \cite {CKS}, where Pad{\'e} approximation
was used to obtain  $\Delta ^{(2)} _{N\!A}$ as 
a function of $\beta$. A comparison of our result
for  $\Delta ^{(2)} _{N\!A}$ with the results
for the similar quantity presented in  \cite {CKS} 
shows, that for $\beta \ll 1$ there is a reasonable
agreement.

The details of the derivation of the above results cannot be presented
here for lack of space.  We only briefly explain how they are obtained
using the asymptotic expansion in the threshold region. The
contributions of all momenta regions, relevant for this calculation,
are separated according to a classification given above
and the integrands of the loop integrations are expanded
in the respective small parameters (which vary from one momentum
region to another).  All divergences which arise in the course of this
procedure are regulated using dimensional regularization.  The
calculation of the hard contribution is done by solving a system of
recurrence relations which allows to reduce any hard integral to a
limited set of master integrals. All integrals, where only potential
lines are involved, are similar in their form to the integrals which
one encounters in the non--relativistic perturbation theory and can be
done easily.  When some lines of a given diagram are either soft or
ultrasoft, the calculation can be performed loop-by--loop.  For this,
the results for the one--loop eikonal integrals \cite {maxtech} are
useful as well as some results obtained in \cite {Peter}.

Let us elucidate the origin of various terms in the above expressions
for $\Delta ^{(2)}_A$ and $\Delta ^{(2)}_{N\!A}$.  The most
interesting contributions come from the hard subgraphs, because these
contributions could be used to discuss the renormalization of the
Sommerfeld--Sakharov factor at the next--to--leading order as well as
a matching of the NRQCD quark--antiquark  vector current on the
full QCD vector current at order ${\cal O}(\alpha _s^2)$ and the
leading order in $1/m$.

We note the following: in both $\Delta^{(2)}_{A,N\!A}$ the terms which
are not accompanied by powers of $\pi$ come from hard subgraphs. These
are $39/4-\zeta_3$ and $-151/36-13\zeta_3/2$, respectively, for
$\Delta^{(2)}_A$ and $\Delta^{(2)}_{N\!A}$. The terms of the order
${\cal O}(\beta^0)$, which are multiplied by a single power of
$\pi^2$, are more tricky.  First, one sees that there is a $\log
\beta$ contribution in these pieces. This implies that these terms get
contributions from both hard and (all possible) soft scales, and these
contributions are not finite separately.  This in turn means that at
the ${\cal O}(\alpha_s^2)$ order the matching coefficient of the NRQCD
vector quark--antiquark current is not finite any longer.  Removing
this divergence by using the $\overline {MS}$ renormalization of the
low energy effective field theory, we arrive at finite (but scale
dependent) matching coefficients.

We write 
\begin{equation}
\bar \psi \gamma ^i \psi  = 
\left ( 1 -2 C_F\frac {\alpha _s}{\pi} 
+ c_2 (\mu) C_F \left (\frac {\alpha_s}{\pi} \right ) ^2 \right )
  \left [\psi _h^{+} \sigma ^i \chi_h \right ]_{\mu},
\end{equation}
where the operator at the LHS of the above equation is the
quark--antiquark current in the full QCD, while the RHS represents the
quark--antiquark current in the effective field theory (with $\psi _h$
and $\chi_h$ being two-component spinors),  multiplied by a Wilson
(matching) coefficient.  We note that to the order 
${\cal O}(\alpha _s, 1/m^2)$ the matching condition for the 
quark--antiquark vector current has been obtained in 
\cite {Luke97}.

The knowledge of contributions coming from hard subgraphs
allows us to obtain this matching coefficient directly: 
\begin{eqnarray}
c_2 (\mu) &=& C_F c_A +C_A c_{N\!A} + N_LT_R c_L +N_H T_R c_H,
\\
c_A (\mu) &=&
\pi ^2 
\left (-\frac {79}{36}-\frac {1}{3} \log\left (\frac {\mu}{m} \right )
 + \log 2
\right )
\nonumber \\&&
+\frac {23}{8} -\frac {1}{2}\zeta _3,
\\
c_{N\!A} (\mu) &=& 
\pi^2 \left (\frac {89}{144}-\frac {1}{2} \log \left (\frac {\mu}{m}
\right) 
 -\frac {5}{6}\log2 \right )
\nonumber \\&&
-\frac {151}{72}-\frac {13}{4}\zeta _3,
\\ 
c_{L} (\mu) &=& \frac {11}{18},
\\ 
c_H(\mu)  &=& -\frac {2}{9}\pi^2+\frac {22}{9}.
\end{eqnarray}

We note finally that within the present approach it is possible to
calculate also the higher order terms in the expansion in
$\beta$. However, already for the terms of order $\beta^1$ in
$\Delta^{(2)}$ one should incorporate the real gluon radiation.  The
expansion of the real radiation and the phase space in this case can
be obtained as a slight generalization of a similar expansion,
discussed in \cite {maxtech} in the context of heavy quark decays.

In conclusion, we have calculated the ${\cal O}(\alpha _s ^2)$
correction to the heavy quark production cross section in $e^+e^-$
annihilation in the threshold region, assuming $\alpha_s \ll \beta \ll
1$.  A method of systematic expansion of the Feynman diagrams near
thresholds in powers and logarithms of the quark velocity $\beta$
enabled for the first time an evaluation of the non-abelian
terms.  At the same time, comparison of the abelian terms with
previously obtained results gives us confidence in the correctness of
the method.  The results of the present paper can be further used for
matching--like calculations, to arrive at the predictions for the
threshold cross section at the energy region, where $\beta \sim
\alpha_s$.  We have also given a matching relation between full QCD
and NRQCD effective currents to leading order in $1/m$ and second
order in the strong coupling constant.

\vspace*{.3cm}

{\bf Acknowledgments:} \hspace{.4em} We are indebted to M.\ Beneke,
K.\ Chetyrkin, V.\ Smirnov and A.\ Yelkhovsky for useful discussions.
We would like to thank Y.\ Sumino for comments regarding this
manuscript. We thank J.~H.~K\"uhn for his interest in this work and
encouragement.  We are grateful to M.~Beneke, A.~Signer, and
V.~Smirnov \cite{BSS} for pointing out an arithmetic error in
eqs.~(12-13) in the first version of this paper.  This work was
supported in part by DOE under grant number DE-AC02-76CH00016, by BMBF
under grant number BMBF-057KA92P, and by Graduiertenkolleg
``Teilchenphysik'' at the University of Karlsruhe.


\begin{thebibliography}{10}

\bibitem{Voloshin:1995sf}
M.~B. Voloshin, Int. J. Mod. Phys. {\bf A10},  2865  (1995).

\bibitem{Bodwin:1995jh}
G.~T. Bodwin, E. Braaten, and G.~P. Lepage, Phys. Rev. {\bf D51},  1125
  (1995).

\bibitem{Hoang1} A.H. Hoang, Phys.Rev. {\bf D56}, 5851 (1997).

\bibitem{Hoang:1997sj}
A.~H. Hoang, Phys. Rev. {\bf D56},  7276  (1997).


\bibitem{CKM96}
A. Czarnecki, B. Krause, and W. Marciano, Phys. Rev. Lett. {\bf 76},  3267
  (1996).

\bibitem{Czarnecki:1997hc}
A. Czarnecki and K. Melnikov, Phys. Rev. Lett. {\bf 78},  3630  (1997).

\bibitem{Chetyrkin:1996ia}
K.~G. Chetyrkin, J.~H. K{\"u}hn, and A. Kwiatkowski, Phys. Rept. {\bf 277},
  189  (1996).

\bibitem{Beneke:1997zp}
M. Beneke and V.~A. Smirnov, hep-ph/9711391 (unpublished).

\bibitem{Smirnov90}
V. Smirnov, Comm. Math. Phys. {\bf 134},  109  (1990).

\bibitem{Chetyrkin:1988zz}
K.~G. Chetyrkin, Theor. Math. Phys. {\bf 75},  346  (1988).

\bibitem{Chetyrkin:1988cu}
K.~G. Chetyrkin, Theor. Math. Phys. {\bf 76},  809  (1988).

\bibitem{Hoang:1995ex}
A.~H. Hoang, J.~H. K{\"u}hn, and T. Teubner, Nucl. Phys. {\bf B452},  173
  (1995).

\bibitem {CKS} K.G. Chetyrkin, J.H. K\"uhn, M.Steinhauser,
 Nucl. Phys. {\bf B482}, 213 (1996).

\bibitem{Luke97} M. Luke and M. Savage, hep-ph/9707313 (unpublished).

\bibitem{maxtech}
A. Czarnecki and K. Melnikov, Phys. Rev. D {\bf 56},  7216  (1997).

\bibitem{Peter}
M. Peter, Nucl. Phys. {\bf B501},  471  (1997).

\bibitem{BSS}
M. Beneke, A. Signer, and V. Smirnov, hep-ph/9712302. 
\end{thebibliography}

\end{document}